\documentclass[a4paper]{article}

\usepackage[margin=1in]{geometry}

\usepackage{amsmath, amssymb}

\usepackage{graphicx}
\usepackage{float}
\usepackage{rotating}
\usepackage{placeins}
\usepackage{longtable}

\usepackage{booktabs}
\usepackage{colortbl}
\usepackage{xcolor}
\definecolor{dj}{RGB}{150,10,220}
\definecolor{ron}{RGB}{150,10,220}

\usepackage{enumitem}

\usepackage{comment}

\usepackage[short, c3]{optidef}

\usepackage[natbibapa]{apacite}
\usepackage[utf8]{inputenc}

\usepackage{hyperref}
\usepackage{url}

\allowdisplaybreaks

\makeatletter
\let\@fnsymbol\@arabic
\makeatother

\title{Maritime Port Supply Chain Resilience: A Systematic Review}

\author{%
  \footnotemark[1] \ Digvijay Redekar\footnotemark[2]\\
  \href{mailto:dredekar@asu.edu}{\texttt{dredekar@asu.edu}}
  \and 
  Ronald Askin\footnotemark[1]\\
  \href{mailto:ron.askin@asu.edu}{\texttt{ron.askin@asu.edu}}
  \and
  Feng Ju\footnotemark[1]\\
  \href{mailto:fengju@asu.edu}{\texttt{fengju@asu.edu}}
}

\date{}

\begin{document}

\maketitle

\footnotetext[1]{
  Industrial Engineering Department, Arizona State University 
}

\footnotetext[2]{
  Corresponding author, \href{https://orcid.org/0009-0003-8541-6661}{ORCID: 0009-0003-8541-6661}
}

\newpage

\begin{abstract}
Maritime transportation systems (MTS) play a crucial role in ensuring the uninterrupted supply of essential goods and services, impacting the economy, border security and general welfare. However, MTS operations face disruptions from natural disasters, man-made disturbances, or cascading combinations of these events. These threats can disrupt trade routes, pose serious risks to personnel, and damage critical port infrastructure. In today’s advanced technological world, MTS also face growing threats from cyberattack and terrorist attack. Given these evolving risks, efforts to ensure resilience in port operations must be more inclusive, reflecting the full spectrum of potential threats. This article focuses on two key aspects of maritime transportation supply chain resilience (MTSCR): operational excellence and system technology. It presents a structured review framework of relevant literature including research papers and government documents, to explore strategies for strengthening MTSCR. The article classifies the review findings and highlights vulnerable areas requiring future research.


\end{abstract}

\section{Introduction}\label{sec1}

Supply chain resilience (SCR) is the ability of supply chain systems to function effectively with minimal impact in the event of disturbances and the ability to respond swiftly to mitigate disruptions in supply chain operations. Historically, humankind has experienced natural and manmade disasters causing supply chain instability, but factors such as globalization, climate change, economic expansion and product technology have exacerbated the likelihood and impact of supply chain disruptions. Supply and network disruptions have resulted in significant economic impacts limiting, at times, access to food and energy supplies and shutting down factory production. Natural disasters can be earthquakes, tsunamis, tornadoes, hurricanes, floods and pandemics. Manmade disruptions include shipping canal and bridge accidents, cyberattack, equipment malfunction, political instability, and similar disastrous events. In some cases, disturbances may have multiple causes, leading to cascading impacts. For example, poor safety protocols and lean inventory can result in the shutdown of baby formula production facilities, potentially leading to an unprecedented infant health crisis and long-term adverse health effects.

Understanding and addressing the vulnerabilities of specific supply chain systems, such as maritime port supply chains, is crucial to achieving resilience in critical sectors of the economy. Maritime ports serve as vital nodes in global trade, handling large volumes of goods and raw materials. However, these systems are increasingly vulnerable to disruptions due to the complexity of modern port operations, reliance on advanced technologies, and interdependencies with other supply chain sectors. Identifying and addressing vulnerabilities in maritime ports is particularly important in the United States, where SCR is closely tied to economic and national security. Disruptions in maritime port operations can have cascading effects on industries dependent on timely imports and exports, further highlighting the need for focused research in this area.

The goal of this research is to perform a literature review of resiliency of maritime port supply chains mainly focused on United States ports and to identify potential vulnerable areas requiring future research that will strengthen port operations and improve resilience. This research focuses mainly on two domains of the port supply chain industry, namely operational excellence and cybersecurity. These domains are susceptible to disruption incidents and pose significant economic and national security threats. The optimal, continuous functionality of MTS is demanded to avoid any delays in supply chain operations. Port operations study focuses on operational performance and risk avoidance policies. Whereas cybersecurity focuses on threats to physical systems and port technology. The study discusses causes of historical disruptions in port operations and cybersecurity, and solutions identified by researchers either implemented or recommended. As a significant volume of goods and materials are imported through ports, even minor disruption in port operations can cause substantial impact to many industries, and even temporary economic instability. Hence the research in these domains focuses on potential causes of disruptions and their impact on MTSCR. We categorize them to understand the types of disruptions, strategies implemented to mitigate the impact, and summarize the results.

\section{Background}\label{sec2}

Several previous reviews of SCR literature have been published. In this section, we summarize some of those literature reviews to the best of our knowledge to understand the latest topics discussed pertaining to SCR in MTS domain. Researchers' specific focus in SCR is discussed along with their findings and conclusion. This provides a foundation for our research, synthesizing and updating the current state of knowledge while serving as a springboard for identifying critical future research directions in the respective topics.

In October 2024, the SCRIPS (Supply Chain Resilience Issues, Problems and Solutions) workshop, sponsored by the Department of Homeland Security Centers of Excellence, convened scholars and government experts to discuss resilience strategies for maritime port, agriculture, and semiconductor supply chains \citep{Askin2024}. Discussions emphasized the need for advanced modeling techniques, such as system mapping and digital twins, to identify vulnerabilities and simulate disruption scenarios \citep{Askin2024}. Key challenges included labor shortages, resistance to automation, and climate risks, necessitating investments in adaptive infrastructure and scenario-based planning. Participants stressed improving data collection, fostering stakeholder collaboration, and integrating AI-driven automation with human behavior principles to enhance operational efficiency and resilience \citep{Askin2024}. \citet{Akpinar2023} conducted a systematic literature review to synthesize existing approaches and proposed a framework tailored to enhance port resilience amidst disruptions. Findings suggest that organizational resilience in this sector is underdeveloped, particularly in terms of integrating diverse resilience practices and frameworks across varying scales and scenarios. The researchers propose a more unified model of resilience that considers the unique vulnerabilities of the maritime industry, including its exposure to global supply chain disruptions, environmental risks, and economic volatilities. Organizational resilience in maritime sector is underdeveloped, particularly in terms of integrating diverse resilience practices and frameworks across varying scales and scenarios \citep{Gu2023}. The conclusion points towards a need for comprehensive strategies that not only address immediate disruptions but also build long-term resilience. \citet{Bosco2019} reviewed the existing literature on the impact of port disruptions on the maritime supply chain. MTS resilience can be improved by considering multiple resiliency actions instead of singular approach. MTSCR is in early stages of the development compared to ground and air transportation. The review pointed out the scarcity of quantitative methodologies considering comprehensive resilience actions, particularly ship rerouting, and the need for more research on port sustainability and resilience to climate change. The conclusion points towards a need for comprehensive strategies that not only address immediate disruptions but also build long-term resilience against a spectrum of potential threats. Developed countries experience the scarcity of research on multimodal transport resilience, particularly at the interface of maritime and land-based transport in seaport settings \citep{Madhusudan2011}. Their review identify areas requiring further attention, such as resilience measures, port-related multimodal systems, and comprehensive disaster studies in countries like India, China, Brazil, and Russia.

Based on the analyzed reviews, we identify a lack of research in the SCR domains related to the focal areas discussed in Section \ref{sec1}. The next section outlines the methodology used for this literature review. The Findings section presents the main ideas from the existing relevant research, while the Conclusion section elaborates on potential research directions.

\section{Methodology}\label{sec3}

\begin{figure}
\centerline{\includegraphics[scale=0.55]{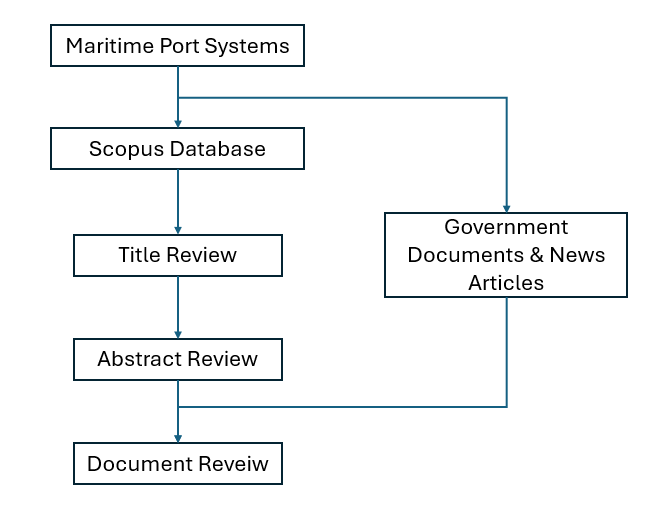}}
\caption{Methodology Steps}
\label{fig:methodology}
\end{figure}

Research was conducted using academic research articles, publicly available reports, and online articles. Academic research articles are sourced from the Scopus database. Publicly available reports and news articles were searched through internet such as google search and mainstream news articles. These articles and reports are published papers in archival journals, public access government reports, academic research project reports, industry published reports or individual researched published reports. We discuss the source and method of research reviews of articles and reports in detail below and the steps of review methodology is shown in \autoref{fig:methodology}. Academic research articles were sourced using Scopus database. We searched the Scopus database source with predefined keywords related to SCR. The Scopus database is a comprehensive online library from academic publisher Elsevier and provides collection of relevant research articles including papers, conferences, reviews, book chapters, etc. Total articles screened for maritime port SCR were 684 and search keywords are shown in \autoref{table:martime_SCR}. The ``Repetition" column provides the frequency of the keyword appearing in title, abstract or article keywords, whereas ``\% Repetition" shows frequency of respective keyword in percentage format. In the ``Search Syntax Prefix” column, we delineate the search logic employed in the Scopus database: core terms related to maritime port supply chain resilience are combined with ``and" to target the paper’s primary objective, while supplementary keywords are linked with ``OR” to broaden the scope and capture relevant, contextually related literature.

We screened 684 articles from the Scopus library based on keyword searches. We utilized filters available on Scopus dashboard to source relevant articles. The filter selection criteria is discussed in \autoref{tab:criteria}. These papers were scrutinized in three steps. The first step was a title review, where we read the titles and associated author keywords of all papers to identify relevance to SCR. This step helped remove unrelated papers from the database. It also provided a relevance strength for each keyword, as denoted in the 'Repetition' column. Relevance strength was calculated based on the frequency of each keyword appearing in paper titles, author keywords, and index keywords. The title review resulted in 193 papers, which were analyzed in the second step, the abstract review. The abstract review involved examining the article’s problem statement, goals, and overall methodology mentioned in the abstract. This step helped in understanding the scope and intent of the research, an overview of the methodology and analysis, and a summary of the findings. The third step involved studying these 73 articles for relevance to our research goals and analyzing their findings. We also utilized internet searches to identify SCR-relevant news articles, government documents, and reports. This process resulted in a final list of 59 articles and reports included in this research. The Findings section provides research summaries, main ideas, methodologies, and conclusions. The Discussion section includes takeaways from these research ideas and identifies potential directions for future research.

\begin{table}
\centering
\begin{tabular}{|l|l|l|l|}
\hline
\textbf{Maritime Port SCR}         & \textbf{\% Repetition} & \textbf{\# Repetition} & \textbf{Search Syntax Prefix} \\ \hline
supply                    & 97.8\%                 & 669                    & and                            \\ \hline
chain                     & 88.5\%                 & 605                    & and                            \\ \hline
port                      & 87.1\%                 & 596                    & and                            \\ \hline
operation                 & 76.2\%                 & 521                    & or                             \\ \hline
logistics                 & 62.7\%                 & 429                    & or                             \\ \hline
maritime                  & 45.5\%                 & 311                    & and                             \\ \hline
security                  & 42.0\%                 & 287                    & or                             \\ \hline
trade                     & 20.8\%                 & 142                    & or                             \\ \hline
attack                    & 17.7\%                 & 121                    & or                             \\ \hline
resilience                & 12.1\%                 & 83                     & and                            \\ \hline
cyber                     & 11.4\%                 & 78                     & or                             \\ \hline
\end{tabular}
\caption{Keywords for Maritime and Logistics Security}
\label{table:martime_SCR}
\end{table}

\begin{table}
\begin{tabular}{|c|c|p{4cm}|p{6cm}|}
\hline
\# Number & Criteria      & Description                                                 & Reasoning                                                                                                                                                                                                                       \\ \hline
1         & Year          & Articles published in the following year range: 2003 to present & Articles from the last 20 years to identify the latest trends, up-to-date methodologies, recent disruptions, and avoids outdated methodologies that may no longer be relevant \\ \hline
2         & Language      & Articles published in English language                       & Based on a common language fully understood by authors, it avoids the risk of misinterpretation in translation and ensures relevancy to the expected audience                              \\ \hline
3         & Document type & Article, review, conference paper, conference review, book, book chapter & Document types that are main source of the relevant studies \\ 
\hline
4 &	Domain	& Maritime port systems, supply chain resilience	& Articles related to maritime port supply chain only\\

\hline

\end{tabular}
\caption{Criteria for Article Selection}
\label{tab:criteria}
\end{table}

\section{Findings}

Resilience in port operations and systems is critical for maintaining the flow of global trade amidst various disruptions. Port operations are vulnerable to a wide range of disturbances, including natural disasters, operational failures, cyber-attacks, and infrastructure damage. Such disruptions, which are unique to port supply chain operations, can have catastrophic consequences for a country's economy and national defense. The impact of port disturbances is more severe compared to regular supply chain disruptions due to their effect on numerous companies relying on seamless first-mile logistics transactions. Major disruptions can lead to delays exceeding a day and cause cascading, indeterminate delays in future operations. These issues also impact incoming goods awaiting offloading, resulting in bottlenecks and increased dwell times. In 2021, the average dwell time for the top 25 U.S. container ports was 32 hours (+3.9 hours compared to 2020) \citep{BTS2023PortPerformance}. Conversely, tanker vessels carrying liquid goods, such as petroleum, experienced a slight reduction in dwell times from 41.4 hours to 40.8 hours \citep{BTS2023PortPerformance}. While increased dwell times for container vessels raise operational concerns, significant disruptions could drastically impact tanker vessel operations, as 40.8 hours is already a prolonged dwell time \citep{BTS2023PortPerformance}. Although modern ports are equipped with disaster management tools and strategies, the impacts of major disruptions remain largely unforeseeable. Such incidents can have multifold consequences, not only for the economy but also for day-to-day life. For example, the COVID-19 pandemic led to the temporary closure of many ports, causing shortages of essential supplies, surge pricing, and, in extreme cases, life-threatening situations. This research aims to analyze port resiliency, assess the current state through case studies, and provide future research directions for improving port operations and systems. Port resiliency involves robust risk assessment and management strategies that address dependencies between information technology and operational technology systems. Based on the literature study, we classify port literature into five main sections as shown below. The first category focuses on protecting IT systems and the integration of emerging technologies, and the second addresses the physical and logistical components essential for port functionality, aimed at enhancing efficiency and resilience. The third section classifies disruptions ranging from natural disasters to manmade threats and operational pitfalls along with strategies to mitigate them. The fourth highlights operational excellence by detailing analytical models and theoretical frameworks that improve performance and resiliency. Lastly, the fifth section reviews recent government policies, emphasizing how legislative and regulatory measures can significantly shape and support port resilience efforts. A robust, integrated framework spanning these five domains fortifies ports against a wide spectrum of potential disruptions and expedites recovery, thereby preserving uninterrupted operations.

\begin{enumerate}
    \item Infrastructure and critical commodities
    \begin{enumerate}
        \item Smart ports and automation
        \item Early warning systems
    \end{enumerate}
    
    \item System technology and cyber security
    
    \item Disruptions classification
    \begin{enumerate}
        \item Natural disruptions
        \item Manmade disruptions
        \item Operational pitfalls
    \end{enumerate}
    
    \item Operational excellence
    \begin{enumerate}
        \item Analytical and simulation models
        \item Theoretical frameworks
    \end{enumerate}

    \item Review of recent government policies
\end{enumerate}

\subsection{Infrastructure and Critical Commodities}

This section addresses concerns related to infrastructure and critical commodities within port supply chains, including port health, capabilities to manage large vessels or heavy traffic, smart port technologies, power and telecommunication systems, navigation systems, spare parts, and container availability. The Cybersecurity and Infrastructure Security Agency (CISA) defines port infrastructure resilience as the ability to prepare for, adapt to, withstand, and rapidly recover from disruptions caused by natural disasters or human-made hazards \citep{CISA2021}. Despite its wealth, the United States ranks 13th globally in infrastructure protection quality, indicating a gap in resilience against physical and cyber threats \citep{Roshanei2021}. Aging infrastructure remains a major concern for many older ports in the U.S. For example, Port Miami, a key player in both cargo and cruise operations, maintains strong regional connections with ports like Freeport, Nassau, and Kingston. However, five of Port Miami's top ten partner ports have been identified as highly vulnerable to climate-related disruptions, posing significant risks to the overall network's resilience \citep{Grant20241}. This underscores the need for all MTS ports to integrate climate risk considerations into their infrastructure upgrades \citep{Grant20241}. U.S. seaports are actively implementing resilience,  enhancing construction and design strategies, such as stormwater management infrastructure improvements, to address climate change challenges \citep{Kalaidjian2022}. Globally, the projected adaptation costs for mitigating risks at ports due to sea-level rise range between \$223 billion and \$768 billion by 2050 \citep{ABDELHAFEZ2021103884}. Ports function as critical nodes in supply chains, where disruptions can cause cascading production interruptions and economic losses across multiple sectors and regions. Analytical tools, such as the enormous regional model (TERM) a multiregional computable general equilibrium (CGE) framework are used to evaluate the resilience of critical port infrastructure. Simulations conducted by \citet{WEI2020} demonstrated that applying inherent resilience tactics, including input substitution, import substitution, and relocation of economic activities, can reduce potential GDP losses from port disruptions by over 90\% \citep{WEI2020}. To address aging infrastructure and climate-induced vulnerabilities, ports worldwide are adopting innovative resilience strategies. For instance, \citet{Becker2017} proposed the “Protect” strategy for Providence Harbor, which involves constructing a hurricane barrier modeled after the Maeslantkering Barrier. This barrier is designed to withstand a 6.4-meter storm surge, safeguarding critical infrastructure, such as petroleum terminals and transportation networks, from extreme weather events \citep{Becker2017}. This example highlights the importance of modernizing port infrastructure to ensure operational continuity under increasing climate risks.

\subsubsection{Smart Ports \& Automation}

Traditional port operations have become increasingly inefficient in comparison to the technological advancements of the 21st century. The concept of a smart port leverages advanced digital and green technologies to enhance operational efficiency, sustainability, and competitiveness within the MTS \citep{Basulo_Ribeiro_2024}. Port operations can be optimized through the integration of smart technologies, such as the Internet of Things (IoT), smart sensors, automated guided vehicles (AGVs), and advanced communication systems \citep{Hayder2023, Koroleva20219}. These technologies not only improve efficiency but also enhance environmental sustainability and boost economic performance \citep{Hayder2023}. Smart port infrastructure integrates three core components automation, environment, and intelligence. Systems such as automated handling, remote control, and renewable energy have been shown to enhance operational performance \citep{YEN2023100862}. \citet{YEN2023100862} conducted quantitative analysis on partial automation, including automatic scanning and handling systems, demonstrating a 12.7\% improvement in operational throughput. Automated machinery, including automated stacking cranes, quay cranes, and integrated infrastructure (e.g., GPS, RFID, and Port Community Systems i.e. PCS), play pivotal role in enabling smart port operations by reducing human errors, alleviating congestion, and improving vessel turnaround times \citep{Molavi_2020, Bastug2020}. The Port of Rotterdam’s digital twin leverages IoT sensors and augmented intelligence to reduce vessel stay times, optimize berth availability, and increase daily vessel handling capacity \citep{Koroleva20219}. Accurate hydro and meteorological data, such as air temperature, wind speed, and water levels, enable cost-effective cargo delivery and safe navigation, contributing to reduced marine fuel consumption and minimized operational delays \citep{Koroleva20219}. The transformation of traditional ports into smart ports relies on two main pillars: inter-connectivity of the port logistics chain and automation \citep{Douaioui2018}. Inter-connectivity enables real-time data exchange among stakeholders to improve coordination, minimize delays, and optimize logistics flows. Automation is critical for port operations, including smart ships, smart containers, and automated handling systems \citep{Douaioui2018}. \citet{Rajabi2018} analyzed the implementation of smart port architecture through the integration of IoT devices, RFID sensors, and cloud/fog computing for real-time monitoring. This approach optimized trade flows and minimized delays, with Automatic Identification Systems (AIS) data from the Port of Le Havre showing that the "Port 2000" terminal received 39.39\% of Ultra Large Container Ships, highlighting traffic imbalances across terminals. IoT-enabled sensing systems, such as RFID, laser, and ultrasonic sensors, play a key role in smart ports by tracking container positions, monitoring infrastructure health, and enhancing automated container handling using AGVs and quayside cranes \citep{Yang_Yongsheng2018, Koroleva20219}. Information and communications technology systems further support smart ports by leveraging IoT platforms and real-time data analytics to streamline operations across terminals, quays, and storage yards. These technologies enable seamless coordination among stakeholders and improve overall port efficiency \citep{Yau_kom2020}. Additionally, smart energy management systems, such as cold ironing (shore-to-ship power supply) and motion-sensitive lighting, reduce emissions and operational costs while improving energy efficiency across port facilities \citep{Yau_kom2020}.

\subsubsection{Early Warning Systems}

As noted previously, aging infrastructure remains a significant concern for many older ports in the United States. Port Miami, for instance, maintains strong connections with key regional ports such as Freeport, Nassau, and Kingston, which are vital for both cargo and cruise operations \citep{Grant20241}. However, five of Port Miami's top ten partner ports have been identified as highly vulnerable to climate-related disruptions, posing considerable risks to the resilience of the overall network \citep{Grant20241}. To address climate change risks, U.S. seaports frequently adopt construction and design strategies, including stormwater management infrastructure improvements, as part of resilience enhancement efforts \citep{Kalaidjian2022}. Globally, the projected adaptation expenses to mitigate port risks due to sea-level rise are estimated to range between $223 billion and $768 billion by 2050 \citep{ABDELHAFEZ2021103884}. The multi-hazard approach proposed by the Intergovernmental Panel on Climate Change (IPCC) under high-emission scenarios provides comprehensive risk assessments for ports worldwide. This methodology is flexible for both global and regional analyses, supporting economic impact evaluations and targeted resilience planning \citep{Izaguirre2021}. Integrating advanced climate prediction models with real-time data analytics in port operations enables contingency strategies to endure climatic extremes and improves MTSCR \citep{Desroches2014649}. The Port Security Grant Program by the Federal Emergency Management Agency provides federal funding to maritime critical infrastructure and key resources, supporting resilience-based approaches for port security. Models that integrate traditional risk assessment with resilience-focused strategies, leveraging network science and operations research, offer quantifiable measures to enable early warnings, prevention, and mitigation \citep{Taquechel2013521554}. Awareness of vulnerabilities among MTS stakeholders is essential for preparedness. \citet{Berle2011605} conducted surveys and interviews to emphasize the critical lack of preparedness for low-frequency, high-impact events, which can severely disrupt the maritime supply chain. The study highlights the importance of metric-based evaluations to identify potential threats and develop robust mitigation plans. Operational improvements, such as the implementation of truck announcement systems (TAS) in container terminals, address stacking inefficiencies and congestion. TAS provides advanced notifications of truck arrivals, allowing terminals to optimize container placement and reduce unproductive reshuffles. Simulations by \citet{Asperen2011} demonstrated that TAS can increase the percentage of containers accessible without reshuffles to over 90\% when announcements are made 4–12 hours in advance. Disaster management strategies ensure the mitigation of impacts and the rapid resumption of normal operations. \citet{Russell200344} proposed secure telecommunication channels and media frameworks to effectively manage crises. Timely communication of accurate information to disaster relief officials is critical for accelerating recovery efforts and minimizing casualties \citep{Russell200344}.

\subsection{System Technology and Cyber Security}

This section evaluates the existing technologies utilized by ports, their vulnerabilities to disruptions such as cyberattack and ransomware, and opportunities for technological improvement through automation, compliance, and information-sharing systems. Technologies deployed at ports play a critical role in securing operations against online threats while enhancing overall efficiency. Wireless communication systems such as ZigBee, Wi-Fi, and RF are fundamental to smart port operations, enabling real-time communication and security, although challenges persist due to interference from large metal equipment and high-energy devices \citep{Hayder2023}. The IoT integrates various sensors and devices to facilitate container surveillance, infrastructure monitoring, and intelligent transportation systems \citep{Hayder2023}. The Tobit regression model by \citet{YEN2023100862} analyzes the relationship between port technologies and operational efficiency, particularly when data values are censored or constrained. Results demonstrated that artificial intelligence improved efficiency by 9.8\%, whereas information-sharing systems reduced efficiency by 16.7\% due to challenges in processing large data volumes, underscoring the need for strategic adjustments in implementing automation and data transfer technologies \citep{YEN2023100862}. \citet{WEAVER2022103423} applied the Nearly-Orthogonal Latin Hypercube (NOLH) experimental design to simulate disruption profiles for ports, facilitating quantitative assessments of cyber-induced economic losses. Additionally, their Dynamic Discretization Discovery (DDD) algorithm models physical impacts and recovery costs of cyber disruptions while emphasizing the dependencies between Information technology and operational technology systems that are critical to port resilience. \citet{Weaver2021192} developed CITE2 Uniform Resource Names (URNs) syntax system for scientific data management and theoretical framework for analyzing risk associated with critical infrastructure asset. Cybersecurity remains a key concern, particularly with systems like Supervisory Control and Data Acquisition (SCADA), which are essential for monitoring and controlling physical operations but remain vulnerable to cyber threats \citep{CISA2021, Roshanei2021}. Between 2009 and 2019, the energy and transportation sectors recorded the highest number of cyber incidents, with 130 global incidents, driven largely by ransomware and wiper malware attacks such as WannaCry and NotPetya \citep{Roshanei2021}. The report by \citet{CISA2021} recommends implementing robust cybersecurity measures alongside dependency analyses to ensure the resilience of critical systems. At the Port of Rotterdam, the integration of PCS, IoT-enabled tracking, and Big Data analytics led to a 25\% reduction in vehicle idle times, a 30\% improvement in yard efficiency, and a 15\% reduction in overall operational costs \citep{Bastug2020}. These technologies facilitate real-time monitoring, data-driven decision-making, and enhanced resource utilization. \citet{Rajabi2018} demonstrated the use of cloud and fog computing for managing large volumes of AIS data, storing over 4,009,864,124 messages across 32 months and occupying 550 GB of storage, showcasing the scalability of the MongoDB NoSQL database for big data management in smart port environments. Scalable microservices-based architectures have also been proposed for streamlining port operations. \citet{Amrou2019} developed a microservice framework for roll-on/roll-off (RoRo) terminals, consisting of four key components: MessageBroker, leveraging Kafka for fault-tolerant data collection; DataCollection, using MongoDB for efficient data storage and retrieval; MiningService, employing ProM 6 for detecting process inefficiencies; and AlertService, generating real-time alerts to address deviations and traffic congestion. Information security and compliance systems play a pivotal role in port operations by ensuring legal obligations are met and facilitating automated data exchanges. The International Maritime Organization (IMO) provides critical regulations for risk avoidance and compliance. For instance, PCS systems aid in Verified Gross Mass (VGM) compliance while reducing administrative burdens \citep{Fedi201929, Bastug2020}. Similarly, AIS and Electronic Chart Display and Information Systems (ECDIS) require rigorous vulnerability assessments to mitigate risks, as shown by \citet{Roberts2019}. Wireless communication technologies, including 4G/5G networks, enable real-time data transmission across ports, supporting remote monitoring, quick decision-making, and efficient operation management \citep{Yang_Yongsheng2018}. Coupled simulation and optimization approaches have also been proposed to minimize disruption costs and model commodity flows through ports \citep{Weaver20182747}. The Wecision decision-support tool, proposed by \citet{Becker2017}, aggregates stakeholder preferences to evaluate resilience strategies for ports. In workshops, the tool identified the “Protect” strategy, which includes constructing hurricane barriers as the most effective solution for minimizing disruptions and ensuring business continuity. Integrating information security systems with SCR frameworks allows maritime stakeholders to monitor the cargo lifecycle, reducing operational risks and potential costs \citep{Mansouri2009}. \citet{Russell200344} proposed a conceptual five-tenet framework for secure communication channels and collaborative disaster mitigation strategies, emphasizing the importance of evaluating overseas trading partners to reduce risks.

\subsection{Disruption classification}

\subsubsection{Natural Disruptions}

Climate change introduces unprecedented challenges, such as rising sea levels, which threaten port watergates and restrict docking capabilities for some vessels. Using network analysis, \citet{Grant20241} examined the impact of climate change on Port Miami and identified that five critical partner ports pose significant risks due to climate-related vulnerabilities. Resilience planning interventions can help ports prepare for such risks. For instance, \citet{Kalaidjian2022} conducted an RPI experiment with ten U.S. seaports, resulting in a 27.4\% increase in resource sharing with external stakeholders and a 26.4\% improvement in internal collaboration post-intervention. The Hurricane Interactive Track Simulator (HITS), coupled with historical data from nine major hurricanes (2012–2019), estimated a 3-day shutdown of Port Houston for a 1 in 10-year hurricane scenario \citep{BALAKRISHNAN202258}. Rising sea levels exacerbate storm surge flooding risks, intensifying hurricane impacts on coastal infrastructure. For example, damages to the Port of Mobile from a Katrina-like hurricane could be nearly seven times higher under extreme sea-level rise projections by the late 21st century \citep{ABDELHAFEZ2021103884}. Long-term resilience strategies, such as spatial planning and collaborative governance, are critical for sustaining infrastructure. The "Room for River" program at the Port of Rotterdam demonstrates effective flood management by balancing urban development with environmental resilience \citep{Hein2020}. Similarly, the Port of Portland's Terminal 6 faces seismic risks that could cause severe damage to container handling infrastructure, reducing its annual throughput capacity (ATC). To mitigate these risks, CISA recommended seismic retrofitting of critical infrastructure, including electrical substations, cranes, and communication systems, to maintain operational resilience \citep{CISA_MTS2023}. Ports in hurricane-prone regions, such as Puerto Rico and the Virgin Islands, face significant vessel traffic disruptions during extreme weather events. A case study by CISA utilized AIS data to map vessel traffic patterns and assess network redundancy, supporting supply chain stability during hurricane disruptions \citep{CISA_MTS2023}. Notably, Superstorm Sandy in 2012 caused the shutdown of the Port of New York and New Jersey for over eight days, exposing vulnerabilities such as extensive flooding, electrical failures, and structural impairments \citep{Izaguirre2021, Desroches2014649}. To address such risks, \citet{Valletta2015342} developed a pilot program with a framework based on five measures to quantify the direct and downstream effects of mitigation strategies. \citet{Mansouri2009} studied natural disaster risks at the Port of Boston, particularly focusing on hurricanes. They proposed a Risk Management-based Decision Analysis (RMDA) framework, emphasizing investments in technological redundancy and infrastructural support as cost-effective resilience strategies compared to managing post-disruption consequences. The COVID-19 pandemic further underscored port vulnerabilities, causing widespread disruptions in the global supply chain. Pandemic-related uncertainties, such as compliance requirements and shifting government policies at various ports, delayed ship arrivals and severely impacted operations \citep{Loose2023328}. To mitigate market shocks and improve container handling efficiency, \citet{Loose2023328} developed a predictive algorithm for robust scheduling and management systems. \citet{Smith202350} identified a strong negative correlation between retail inventory indices and import TEU (Twenty-foot Equivalent Unit) indices during the pandemic, with a 4-month lag, indicating that import surges closely followed drops in retail inventories. Additionally, the pandemic, coupled with the U.S.-China trade war, forced route adjustments due to bans on goods like steel and soybeans, while carriers such as Maersk implemented capacity reductions to adapt to market constraints \citep{LI2021100038}.

\subsubsection{Manmade Disruptions}

\citet{UNTD_report2024} examined the impact of geopolitical conflicts in the Black Sea and Red Sea regions, which have led to the rerouting of key shipping lanes. These disruptions have resulted in increased costs and significant logistical challenges, underscoring the need for enhanced diplomatic frameworks and adaptive logistics strategies to mitigate cascading effects on maritime operations. cyberattack are a major source of manmade disruptions, with critical infrastructure becoming increasingly vulnerable. In May 2021, the Colonial Pipeline ransomware attack caused a six-day shutdown, disrupting fuel supplies along the U.S. East Coast and leading to economic losses estimated in the billions \citep{CISA_MTS2023}. To address such vulnerabilities, CISA devised multimodal connections to ensure fuel distribution continuity while implementing infrastructure hardening strategies \citep{CISA_MTS2023}. Similarly, in June 2017, Maersk, a leading shipping and logistics company, suffered a cyberattack from the NotPetya ransomware, resulting in a 10-day operational outage and estimated losses of \$200 million due to inoperable systems across multiple ports. \citet{WEAVER2022103423} proposed a DDD algorithm to simulate disruption profiles and dynamically reconfigure network allocations to minimize costs, enabling economic loss estimation and operational recovery during such cyber-physical disruptions. Between 2009 and 2019, 130 global cyber incidents targeted critical infrastructure, with the energy and transportation sectors being the most affected by attacks from state and non-state actors \citep{Roshanei2021}. \citet{Roberts2019} identified potential combined threats from physical attacks and cyber vulnerabilities, including fake news, cyberattack on operating systems, and disruptions to navigation systems, port gates, and autonomous vehicles. \citet{Thekdi201636} developed the Dynamic Inoperability Input-Output Model (DIIM) to evaluate disruptive scenarios, including dockworker strikes, hurricanes, terrorist attacks, and minor disruptions as baseline cases. Similarly, \citet{Taquechel2013521554} proposed a risk-based quantifiable model to assist the Port Security Grant Program in identifying vulnerable sectors and allocating federal funds to improve resilience against terrorist attacks. The terrorist attacks on September 11, 2001, which targeted the World Trade Center, brought significant attention to enhancing the resilience of critical infrastructure. In response, the United States introduced critical measures such as the 96-hour and 24-hour advance notification rules, the Maritime Transportation Security Act of 2002, and the Customs-Trade Partnership Against Terrorism (C-TPAT) program \citep{Thibault20065}. These initiatives fostered cooperative security relationships between government agencies and private businesses, providing benefits such as reduced inspections and faster processing times at ports for participating firms \citep{Thibault20065}. To improve preparedness, \citet{Russell200344} proposed a five-tenet framework for evaluating and enhancing security-aware logistics and supply chain operations. Their study emphasized the importance of collaboration between industries and government agencies to mitigate terrorist threats and develop comprehensive resiliency plans. \citet{Monson2007} demonstrated the application of network science to model the MTS, identifying vulnerable areas and using disaster simulation models to design effective mitigation strategies.

\subsubsection{Operational Pitfalls}

\citet{UNTD_report2024} highlights operational challenges, including congestion at critical chokepoints such as the Panama Canal, where reduced water levels and rerouted traffic have led to increased delays and rising operational costs. The report also notes a global shortage of skilled seafarers, which complicates the management of digitalized port operations and advanced shipping technologies \citep{UNTD_report2024}. Real-time technologies such as Vessel Traffic Information Systems (VTIS), Automated Gate Systems (AGS), and vehicle emissions monitoring tools have proven effective in reducing traffic congestion and pollution \citep{Basulo_Ribeiro_2024, Loose2023328} addressed uncertainties in container arrival and departure times, which complicate stacking and handling processes. Their reinforcement learning-based algorithm effectively maintained empty stacks for reshuffling and managed new container arrivals, thereby improving SCR \citep{Loose2023328}. Integrated monitoring systems equipped with IoT-enabled sensors, real-time data analytics, and environmental management systems compliant with ISO 14001 standards are critical for mitigating disaster risks. These systems enable monitoring vessel traffic, predicting extreme weather events, and ensuring rapid response to disruptions \citep{Molavi_2020}. For example, the Port of Rotterdam's digital twin, spanning 42 km, integrates real-time data on ship traffic, infrastructure, weather, and geographic conditions to optimize decision-making processes and enable the adoption of autonomous cargo ships by 2025 \citep{Koroleva20219}. Technological innovations such as Auto ID technology architecture which incorporates barcodes, QR codes, and magnetic ID cards for vehicle identification are deployed at terminal checkpoints. Combined with portals that log entry and exit times, this system calculates processing and queue times, reducing manual effort, improving traffic flow, and minimizing check-in delays \citep{Amrou2019}. The Port of Virginia, handling over 2 million TEUs annually, implemented optimal vessel berthing scheduling methods to manage operations across three container terminals. \citet{Thorisson2019} found that high utilization at the least expensive terminal increased the likelihood of delays, reaching 17.5 hours per week. By redistributing vessels to alternate terminals with higher handling costs, delays were minimized, balancing operational expenses and efficiency \citep{Thorisson2019}. \citet{Weaver20182747} analyzed traffic congestion caused by labor shortages and suboptimal gate operations, which led to temporary blockages of optimal routes. They proposed a multicommodity optimal network flow model to identify alternative routes for commodities and adjust capacity to manage disruptions effectively. \citet{Berle2011605} suggested a functional failure mode approach to enhance risk management strategies within the MTS, ensuring ports maintain throughput and support global trade during severe disruptions. Interviews and surveys conducted with MTS stakeholders revealed a lack of awareness regarding existing vulnerabilities, emphasizing the need to prioritize strategy implementations for operational resilience \citep{Berle2011605}. Uncoordinated truck arrivals represent a major source of operational disruptions, leading to increased container reshuffles and extended exit times. TAS address this issue by enabling pre-emptive housekeeping moves during off-peak hours, reducing the workload of automated stacking cranes during peak periods. Experimental results by \citet{Asperen2011} demonstrated that TAS reduced reshuffle rates by 15–25\% and exit times by up to 0.12 hours compared to non-TAS scenarios.

\subsection{Operational Excellence}

This section describes analytical models (mathematical, simulation, case study, etc.), theoretical frameworks and best practices to optimize MTSSR, provides benefit analysis and strategic policies for better decision making. 

\subsubsection{Analytical and Simulation Models}

The MuZero reinforcement learning algorithm has been applied to optimize container stacking procedures by minimizing the number of container movements, establishing policies to assist human operators and defining performance metrics for evaluating stacking efficiency \citep{Loose2023328}. Multimodal transport network modeling approaches identify alternative distribution routes, such as waterborne and rail connections, and assess redundancy effectiveness for mitigating supply chain vulnerabilities during disasters like pipeline cyberattack or infrastructure failures, ensuring critical supply flows and minimizing economic losses \citep{CISA_MTS2023}. For seismic resilience, \citet{CISA_MTS2023} developed a probabilistic network model to evaluate measures like retrofitting electrical systems and reinforcing cranes to maintain throughput capacity, with AIS-based vessel traffic mapping aiding in connectivity analysis and resilience planning. \citet{YEN2023100862} introduced a three-step DEA-Tobit approach that combines Data Envelopment Analysis (DEA) for quantitative metrics and the analytic hierarchy process for qualitative smart port indicators. Ports adopting high levels of automation and environmental controls achieved efficiency scores exceeding 80\%, with top performers predominantly located in Asia. The International Trade Inoperability Input–Output Model (IT-IIM) by \citet{BALAKRISHNAN202258} quantified the economic impacts of port shutdowns, estimating single-day losses for Port Houston between \$554.3 million and \$706.5 million and rising to \$1.96 billion for a 1 in 10-year hurricane and \$4.68 billion for a 1 in 1000-year hurricane. \citet{LI2021100038} developed a global cargo shipping network model integrating 161 ports across 52 countries for container and bulk cargo flows. They applied a Mixed-Integer Linear Programming (MILP) model combined with a Relax-and-Fix Decomposition heuristic and Gradient Descent technique, ensuring accurate traffic pattern estimations. Similarly,  \citet{Vanye2021} proposed a scenario-based preference model to prioritize security initiatives for embedded smart devices, emphasizing continuous updates and stakeholder collaboration. \citet{Molavi_2020} developed the Smart Port Index (SPI) using normalization and weighted aggregation methods to benchmark KPIs for operations, energy, environment, and security, providing a performance evaluation framework. \citet{Bastug2020} proposed integrating big data analytics with Terminal Operating Systems (TOS) to optimize port operations. Their tool analyzed historical and real-time cargo movements, equipment usage, and traffic patterns to predict container stacking needs and optimize resource allocation. \citet{WEI2020} developed the TERM multi-regional CGE model to evaluate port infrastructure resilience, showing that inherent strategies such as input and import substitution reduced GDP losses by over 90\%, while adaptive strategies like ship rerouting and production recapture reduced business interruption losses by 30\%. \citet{Thorisson2019} used a generalized assignment model with Monte Carlo simulations to optimize vessel berthing schedules, reducing delays to as little as 5.2 hours per week at higher operational costs, while cost minimization strategies resulted in delays of up to 17.5 hours per week, highlighting the trade-off between delays and expenses. \citet{Koroleva20219} developed a predictive data analytics system using wind speed, water levels, tides, and currents to forecast optimal port entry times, reducing delays, fuel consumption, and improving resource allocation efficiency. \citet{Amrou2019} applied the ProM 6 process mining toolset to extract insights from event logs, optimize workflows, detect anomalies, and identify bottlenecks, improving traffic flow and operational efficiency at RoRo terminals. \citet{Becker2017} utilized the SLOSH model to simulate a Category 3 hurricane with a 60-year return period, identifying risks such as bulkhead failures and hazardous spills. The Wecision decision-support tool aggregated stakeholder input to evaluate resilience strategies, with the “Protect” strategy emerging as the most effective for minimizing hurricane damages and ensuring operational continuity. \citet{Thekdi201636} applied the DIIM method to simulate economic losses caused by dockworker strikes, hurricanes, and terrorist attacks. \citet{Woo2013} used a Structural Equation Model (SEM) based on data from terminal operators, shipping companies, and freight forwarders to demonstrate the positive influence of supply chain integration on port performance. \citet{Taquechel2013521554} proposed a risk evaluation model integrating network science with probabilistic risk analysis to assess vulnerabilities in critical infrastructure during disruptions such as terrorist attacks and cyber incidents. \citet{Asperen2011} introduced multiple stacking algorithms, including Leveling with Departure Times (LDT), to reduce reshuffles and crane workloads. Using TAS, their experiments showed that direct access retrieval time improved from 78\% to 96.9\%, significantly enhancing container retrieval efficiency. \citet{Monson2007} developed a constrained transportation optimization model to assess disaster impacts, incorporating sensitivity analysis for critical port shutdowns. This model provided contingency plans and strategic policies for improving SCR within MTS.

\subsubsection{Theoretical Frameworks}

\citet{Grant20241} developed a network analysis framework based on connectivity indices and sensitivity assessments of partnering ports. This framework identifies critical partners and prioritizes their vulnerabilities, enhancing strategies for improving SCR. Future supply chain surges can be predicted by analyzing the relationship between consumer demand, retail inventories, and import volumes. \citet{Smith202350} demonstrated this approach by integrating economic indicators beyond traditional transportation data, establishing a robust link between retail inventory levels and import demand. The hierarchical AHP framework proposed by \citet{YEN2023100862} categorizes smart port indicators into automation, environment, and intelligence aspects, enabling stakeholders to prioritize high-impact areas. Indicators like automatic handling systems (0.375 weight) and renewable energy adoption (0.296 weight) were identified as top contributors to improving operational efficiency. Similarly, \citet{Kalaidjian2022} classified resilience planning approaches into three categories: contractor-led vulnerability assessments, FEMA guided Hazard Mitigation Plans, and self-assessment using the Ports Resilience Index (PRI). Their study demonstrated the positive impact of resilience interventions on fostering collaboration and strengthening infrastructure resilience. The IPCC risk modeling framework integrates hazard, exposure, and vulnerability functions to evaluate infrastructure risks globally. \citet{CISA2021} developed a six step assessment methodology involving partner engagement, problem identification, assessment design, data collection, analysis, and action promotion. This methodology incorporates analytical tools such as geospatial mapping, failure analysis, and decision-support frameworks to ensure actionable outcomes. \citet{Roshanei2021} analyzed global frameworks like the National Institute of Standards and Technology (NIST) cybersecurity framework, emphasizing five core functions identify, protect, detect, respond, and recover as essential for securing critical infrastructure. \citet{Izaguirre2021} applied this framework across 2,013 ports worldwide, integrating exposure and vulnerability analyses to assess future risks under high-emission scenarios. \citet{ABDELHAFEZ2021103884} proposed a risk-informed framework combining multi-hazard assessment, fault tree analysis, and simulation-based scenario analysis to enhance resilience. Their results showed that extreme sea level rise and hurricane scenarios could reduce port functionality by up to 65\%, triple recovery times, and significantly increase damage costs. Targeted infrastructure reinforcements and adaptive planning strategies mitigated these risks. \citet{Fedi201929} introduced a systematic framework for integrating PCS, centralizing data exchange among shippers, carriers, and terminal operators to improve logistics security and operational efficiency. \citet{Douaioui2018} highlighted that smart ports, integrated with real-time data analytics and smart information systems, enable operations monitoring, delay predictions, and decision-making enhancements, leading to improved efficiency and resilience. \citet{Rajabi2018} proposed a layered IoT framework integrating AIS data as a core component to support real time decision making for port monitoring, navigation, and emergency management. Their framework processed over 10 million AIS messages per file, demonstrating its capacity to handle extensive data streams and provide actionable insights for smart port operations. \citet{Valletta2015342} designed an evaluation framework based on robustness, resourcefulness, recovery, redundancy, and cascading effects, aiding in assessing hypercritical assets and their impact on MTSCR. Updates to design guidelines incorporating advanced climate projections and asset criticality considerations enhance infrastructure resilience. \citet{Desroches2014649} emphasized robust flood protection measures and safe failure mechanisms, such as spare capacity, to ensure resilience against climatic extremes. \citet{Berle2011605} identified six primary failure modes loss of capacity to supply, financial flows, transportation, communication, internal operations, and human resources forming the basis of an evaluation framework for identifying vulnerabilities and prioritizing mitigation strategies. \citet{Mansouri2009} developed the RMDA framework, consisting of a three-phase process: assessing vulnerabilities, devising resilience strategies, and valuing investment options. This systematic approach supports resilience evaluation and provides a cost-effective decision-making framework for MTS.

\subsection{Review of Recent Government Policies}

On February 21, 2024, U.S. President Joseph Biden issued \citet{ExecutiveOrder114116}, titled "Executive Order on Amending Regulations Relating to the Safeguarding of Vessels, Harbors, Ports, and Waterfront Facilities of the United States". The amendments empowered the Captain of the Port to restrict access to vessels and facilities linked to digital infrastructure posing security risks. The order established security zones with inspection authority and broadened the definitions of "damage" and "cyber incident" to include malicious cyber activities. Furthermore, it mandated stricter conditions for Coast Guard Port Security Cards, emphasizing ownership responsibility for implementing robust security measures. The Commandant of the Coast Guard was tasked with coordinating with the Department of Justice to address escalating cyber threats targeting U.S. maritime infrastructure. \citet{Sparkman2024} highlighted concerns regarding Chinese-made port cranes, particularly those manufactured by Shanghai Zhenhua Heavy Industries Co. (ZPMC). These cranes, capable of remote control and monitoring, raised fears of espionage and compromised port operations. In response, the Biden administration announced a \$20 billion investment under the 2021 Infrastructure Law to strengthen port security and bolster domestic crane manufacturing. Supported by the American Association of Port Authorities, the CRANES Act of 2023 aimed to reduce reliance on Chinese cranes and enhance national security. Similarly, \citet{HR31692023} addressed risks from foreign cranes by mandating inspections of cranes tied to adversarial nations and banning foreign software usage within five years. These measures sought to safeguard military and commercial supply chains from potential cyber threats and operational disruptions.

Singapore’s government implemented targeted policies to enhance port SCR, including the Enterprise Financing Scheme-Green, which provided 70\% risk-sharing funding for developing green technologies like electric vehicles and solar panels, valid until March 31, 2024. The Maritime and Port Authority set a target to attract \$20 billion (US \$14.8 billion) in business spending commitments from maritime companies by 2024 \citep{EFS_Green}. The development of Tuas Port was a cornerstone of Singapore’s port resilience strategy, with a planned handling capacity of 65 million TEUs annually by the 2040s \citep{Tuas_port}. Policies supporting the port’s development included large-scale land reclamation of 648 hectares and the installation of 448 caissons to create 17.7 km of seawall for accommodating larger vessels. To enhance operational efficiency, government-backed initiatives deployed 5G-enabled AGVs, automated cranes, and the Next Generation Vessel Traffic Management System. Sustainability policies targeted net-zero emissions by 2050, supported by incentives for 50\% less carbon-intensive AGVs and the construction of Green Mark Platinum-certified super low-energy buildings \citep{Tuas_port}. 

China’s port SCR efforts focused on modernizing port infrastructure and promoting foreign direct investment (FDI). The Foreign Investment Law (2019) standardized regulations for inbound investments in key sectors, while continuous updates to the Negative List for Market Access progressively removed restrictions in industries such as oil and gas, telecommunications, and financial services \citep{china_port}. These measures encouraged FDI critical for integrating advanced technologies like IoT, 5G networks, and automation, which improved cargo handling efficiency and reduced bottlenecks. Additionally, cybersecurity measures outlined in Guideline No. 25 reinforced the operational stability of port systems, ensuring uninterrupted supply chain operations \citep{china_port}.

The UK's Border Target Operating Model, effective January 2024, introduced phased controls for imports, including risk-based Sanitary and Phytosanitary (SPS) measures to protect biosecurity and public health. Key features included streamlined health certifications, digitization via the UK Single Trade Window, and a focus on high-risk goods. These policies enhanced port SCR by minimizing delays, ensuring safety, and aligning with global trade standards \citep{UK_BTOM}. Similarly, the UK’s Import Control System 2 (ICS2) Release 3, implemented in June 2024, enhanced port resilience by improving pre-arrival security and customs procedures. The system required economic operators to submit Entry Summary Declarations (ENS) electronically, enabling advanced risk assessments and streamlined customs clearance. By integrating real-time data analytics and automated decision-making, ICS2 mitigated delays, enhanced border security, and ensured the continuity of port operations, aligning with broader EU Union Customs Code objectives \citep{ICS2}.

\citet{UNTD_report2024} underscored the strategic importance of maritime chokepoints, such as the Suez Canal and Panama Canal, in global trade. Reduced water levels in the Panama Canal led to 31\% longer shipping distances on affected routes, resulting in a 25\% drop in cargo volumes and a 1\% increase in ton-mile demand. These findings highlighted the urgent need for investments in resilient port infrastructure to adapt to environmental challenges and maintain supply chain efficiency. CISA emphasized the role of resilient infrastructure systems, including energy, water, transportation, and communication networks, in maintaining regional functionality during disruptions \citep{CISA2021, Roshanei2021}. \citet{Thibault20065} identified significant benefits from U.S. government security initiatives, such as improved collaboration between industry and government, while also highlighting research gaps, including funding management for security programs and supply chain recovery strategies following disruptive events. \citet{IMO_mandate_VGM_2016} mandated that, as of July 1, 2016, containers' VGM encompassing the combined weight of the packed container, cargo, packaging, and the container itself must be declared prior to vessel loading. This requirement, part of the Safety of Life at Sea (SOLAS) convention, played a critical role in enhancing maritime and port safety by ensuring proper weight distribution, reducing risks of overloading, and preventing ship instability \citep{IMO_mandate_VGM_2016}.

\section{Discussion}  

Resilience in port supply chain systems has emerged as a critical area of focus, requiring comprehensive strategies that address infrastructure, technology, disruptions, and operational inefficiencies. Ports serve as vital nodes in global trade, yet aging infrastructure and rising climate risks create significant vulnerabilities. \citet{Grant20241} and \citet{Kalaidjian2022} emphasize the importance of enhancing port infrastructure by integrating climate risk considerations and fostering collaboration. The adoption of smart ports and automation technologies offers substantial improvements in operational efficiency. Systems such as IoT-enabled sensors, AGVs, and predictive analytics enable ports to optimize resource utilization, reduce congestion, and improve vessel turnaround times \citep{Molavi_2020, Koroleva20219}. Furthermore, early warning systems, such as predictive data analytics and simulation models, enhance ports' ability to anticipate disruptions caused by environmental extremes like hurricanes and sea-level rise, ensuring better preparedness and quicker recovery \citep{BALAKRISHNAN202258, Koroleva20219}. Information sharing, both voluntary and mandated, allows for improved operational efficiency through better planning. Automated data-sharing platforms like PCS enhance port security while improving operational transparency \citep{Fedi201929, Bastug2020}. Operational pitfalls, such as traffic congestion and container stacking inefficiencies, can be alleviated through predictive models, TAS, and digital twins for port management \citep{Loose2023328, Asperen2011, Koroleva20219}. System redundancy provides another option for enhancing resilience and operational performance.

Technological advancements are essential for addressing system vulnerabilities, particularly cyber threats. Cybersecurity remains a major challenge, with attacks like the NotPetya ransomware and incidents involving Chinese-made cranes exposing weaknesses in digital infrastructure \citep{Roshanei2021, Roberts2019, Sparkman2024}. Solutions such as the NIST Cybersecurity Framework and secure communication protocols emphasize the integration of risk-based strategies to protect ports from cyber-physical disruptions \citep{CISA2021}.   

Disruptions impacting port operations are broadly categorized into natural disasters, manmade events, and operational pitfalls. Natural disruptions, such as hurricanes and rising sea levels, can severely damage infrastructure and cause significant economic losses, as seen in simulations for the Port of Mobile and Port Houston \citep{ABDELHAFEZ2021103884, BALAKRISHNAN202258}. Collaborative governance and resilience planning frameworks, such as those developed by \citet{Kalaidjian2022} and IPCC hazard models \citep{Izaguirre2021}, offer actionable strategies to mitigate these risks. Manmade disruptions, including cyberattack and geopolitical conflicts, exacerbate delays and financial losses. The CRANES Act of 2023 and the Port Crane Security and Inspection Act aim to address security concerns associated with foreign-manufactured port infrastructure, highlighting the role of government intervention in ensuring resilience \citep{HR31692023, Sparkman2024}.  

Optimizing port operations requires robust analytical and simulation models to evaluate performance, predict disruptions, and prioritize mitigation strategies. Analytical tools such as reinforcement learning algorithms, generalized assignment models, and the TERM CGE model provide insights into economic impacts, vessel scheduling, and port resilience strategies \citep{Loose2023328, Thorisson2019, WEI2020}. Similarly, theoretical frameworks, including the AHP and network analysis models, support decision-making by prioritizing resilience indicators, such as automation and environmental sustainability, to enhance port efficiency \citep{YEN2023100862, Grant20241}. Frameworks like the RMDA and DIIM offer structured approaches for assessing vulnerabilities, simulating disruption scenarios, and valuing investment strategies, ensuring effective resilience planning \citep{Mansouri2009, Thekdi201636}. Recent government policies have further emphasized the need for safeguarding port infrastructure. The Executive Order 114116 empowers port authorities to mitigate cyber threats by restricting access to insecure digital infrastructure and implementing inspection protocols \citep{ExecutiveOrder114116}. Policies like the VGM mandate under the SOLAS convention ensure operational safety by addressing cargo weight discrepancies \citep{IMO_mandate_VGM_2016}. Additionally, investments in port infrastructure and domestic manufacturing initiatives highlight the importance of adaptive strategies to strengthen national security and SCR.  

Overall, improving the resilience of port supply chains requires a multi-faceted approach that integrates infrastructure enhancements, advanced technologies, disruption management strategies, and robust operational frameworks. Coordinated efforts between industry stakeholders, policymakers, and technological innovators are essential to address vulnerabilities and ensure the seamless functioning of ports amidst evolving challenges. Classification of literature by domain and methodology specific discussed framework is demonstrated in \autoref{tab:classification} and is sorted by the type of research (i.e., government policy, qualitative and quantitative).

\section{Conclusion}

This literature review highlights the critical role of resilience in maritime port supply chains, emphasizing the need for robust infrastructure, advanced technologies, and comprehensive disruption management strategies. Ports, as critical nodes in global trade, face vulnerabilities from natural disasters, manmade disruptions, and operational inefficiencies, necessitating adaptive planning and integrated resilience frameworks. Key advancements, such as IoT-enabled systems, predictive analytics, digital twins, and TAS have demonstrated potential in optimizing resource allocation, enhancing throughput capacity, and mitigating risks. Meanwhile, cybersecurity frameworks like NIST and PCS integration have become essential for safeguarding ports against escalating cyber threats.

However, several research gaps remain unaddressed. While studies have explored resilience strategies for physical infrastructure and digital systems, there is limited research on the integration of human behavior in automated port systems and its impact on operational decision-making. Similarly, the economic interdependencies between major ports and regional supply chains amidst cascading disruption scenarios require deeper investigation. Although predictive frameworks and simulation models have been developed, their scalability and applicability to small and medium-sized ports need to be validated. Future research should focus on multi-hazard risk assessment methods that integrate climate projections, cybersecurity vulnerabilities, and manmade disruptions into a unified resilience framework. Additionally, the role of artificial intelligence and machine learning in enhancing predictive accuracy of disruption forecasting and recovery planning offers promising avenues for further exploration. The alignment of these research directions with government policies is crucial for strengthening resilient maritime infrastructure. Initiatives such as Executive Order 114116, the CRANES Act of 2023, SOLAS mandates and investments in domestic manufacturing highlight the growing emphasis on safeguarding critical infrastructure from cyber and geopolitical threats, and underscore the importance of operational safety and risk mitigation. By integrating these policies with advanced analytical models, technological innovations, and collaborative governance, port supply chain stakeholders can address current vulnerabilities and prepare for future challenges. Moving forward, fostering partnerships between industry stakeholders, policymakers, and researchers will be instrumental in developing holistic, data-driven strategies to ensure the resilience and sustainability of maritime port supply chains.

\begin{sidewaystable}[]
\footnotesize
\begin{tabular}{c|l|c|c|c| *{7}{c|}}
\toprule
Index & Reference & Type & Policy & Technology & \multicolumn{2}{c|}{Infrastructure} & \multicolumn{3}{c|}{Disruptions} & \multicolumn{2}{c|}{Excellency Methods} \\
\cmidrule(l){4-12} 
& &  &  & & 
\multicolumn{1}{p{1.3cm}|}{Smartport} &  
\multicolumn{1}{p{1.6cm}|}{Commodity} & 
\multicolumn{1}{p{0.95 cm}|}{Natural} &
\multicolumn{1}{p{1.3cm}|}{Manmade} &  
\multicolumn{1}{p{1.45 cm}|}{Operations} &
\multicolumn{1}{p{1.3 cm}|}{Analytical} &
\multicolumn{1}{p{1.45cm}|}{Theoretical} \\
\midrule

1 & UN Maritime Report, 2024 & Gov policy\dag & Y &  &  &  &  & Y &  &  &  \\ \hline 
2 & CISA Maritime Report, 2021 & Gov policy & Y & Y &  & Y &  &  &  &  & Y \\ \hline 
3 & Enterprise Singapore Report, 2021 & Gov policy & Y & Y & Y &  &  &  &  &  & Y \\ \hline 
4 & Maritime \& Port Authority, Singapore 2022 & Gov policy & Y &  & Y &  &  &  & Y &  &  \\ \hline 
5 & Asia Society Policy Institute Report, 2024 & Gov policy & Y & Y & Y &  &  &  &  &  & Y \\ \hline 
6 & UK Cabinet Office, 2023 & Gov policy & Y &  &  &  &  &  &  & Y & Y \\ \hline 
7 & European Commission, 2024 & Gov policy & Y & Y &  &  &  &  &  &  & Y \\ \hline 
8 & Akpinar et al., 2023 & Qualitative &  & Y &  & Y & Y & Y &  &  & Y \\ \hline 
9 & Grant et al, 2024 & Qualitative &  &  &  & Y & Y &  &  &  & Y \\ \hline 
10 & Kalaidjian et al., 2022 & Qualitative &  &  &  & Y & Y &  &  &  & Y \\ \hline 
11 & Abdelhafez et al., 2021 & Qualitative &  &  &  & Y & Y &  &  &  & Y \\ \hline 
12 & Al-Fatlwai and Motlak, 2023 & Qualitative &  & Y & Y &  &  &  &  &  &  \\ \hline 
13 & Douaioui et al., 2018 & Qualitative &  &  & Y &  &  &  &  &  & Y \\ \hline 
14 & Yang et al., 2018 & Qualitative &  & Y & Y &  &  &  &  &  &  \\ \hline 
15 & Yau et al., 2020 & Qualitative &  &  & Y &  &  &  &  &  & Y \\ \hline 
16 & Izaguirre et al., 2021 & Qualitative &  &  &  & Y & Y &  &  &  & Y \\ \hline 
17 & DesRoches and Murrell, 2014 & Qualitative &  &  &  & Y & Y &  &  &  & Y \\ \hline 
18 & Berle et al., 2011 & Qualitative &  &  &  & Y &  &  & Y &  & Y \\ \hline 
19 & Russell and Saldanha, 2003 & Qualitative &  & Y &  & Y &  & Y &  &  &  \\ \hline 
20 & Weaver et al., 2022 & Qualitative &  & Y &  &  &  &  &  &  & Y \\ \hline 
21 & Weaver, 2021 & Qualitative &  & Y &  &  &  & Y &  &  &  \\ \hline 
22 & Fedi et al., 2019 & Qualitative &  & Y &  &  &  &  &  &  & Y \\ \hline 
23 & Roberts et al., 2019 & Qualitative &  & Y &  &  &  & Y &  &  &  \\ \hline 
24 & Mansouri et al., 2009 & Qualitative &  & Y &  &  & Y &  &  &  & Y \\ \hline 
25 & Executive Order 114116,  2024 & Qualitative &  & Y &  &  &  &  &  &  &  \\ \hline 
26 & Act HR 31692023 & Qualitative &  & Y &  &  &  &  &  &  &  \\ \hline 
27 & Thibault et al., 2006 & Qualitative & Y & Y &  &  &  & Y &  &  &  \\ \hline 
28 & IMO VGM, 2016 & Qualitative &  & Y &  &  &  &  &  &  &  \\ \hline 
29 &  Valletta et al., 2015 & Qualitative &  &  &  &  & Y &  &  &  & Y \\ \hline 
30 & Smith et al., 2023 & Qualitative &  &  &  &  & Y &  &  &  & Y \\ \hline 
31 & Basulo-Ribeiro et al., 2024 & Qualitative &  &  & Y &  &  &  & Y &  &  \\ \hline 
32 & Rajabi et al., 2018 & Qualitative &  & Y & Y &  &  &  &  &  & Y \\ \hline 
33 & Hein and Schubert, 2021 & Qualitative &  &  &  &  & Y &  &  &  &  \\ \hline 
34 & Roshanaei, 2021 & Qualitative & Y & Y & Y &  &  & Y &  &  & Y \\ \hline 
35 & Gu and Liu, 2023 & Qualitative &  &  &  &  &  &  &  &  & Y \\ \hline 
36 & Hu et al., 2023 & Qualitative &  &  &  &  &  &  & Y &  & Y \\ \hline
PTO & \dag Government policy \\

\bottomrule
\end{tabular}
\caption{Literature Classification}
\label{tab:classification}
\end{sidewaystable}

\begin{sidewaystable}[]
\footnotesize
\begin{tabular}{c|l|c|c|c| *{7}{c|}}
\toprule
Index & Reference & Type & Policy & Technology & \multicolumn{2}{c|}{Infrastructure} & \multicolumn{3}{c|}{Disruptions} & \multicolumn{2}{c|}{Excellency Methods} \\
\cmidrule(l){4-12} 
& &  &  & & 
\multicolumn{1}{p{1.3cm}|}{Smartport} &  
\multicolumn{1}{p{1.6cm}|}{Commodity} & 
\multicolumn{1}{p{0.95 cm}|}{Natural} &
\multicolumn{1}{p{1.3cm}|}{Manmade} &  
\multicolumn{1}{p{1.45 cm}|}{Operations} &
\multicolumn{1}{p{1.3 cm}|}{Analytical} &
\multicolumn{1}{p{1.45cm}|}{Theoretical} \\
\midrule

37 & Bosco et al., 2019 & Quantitative &  & Y &  & Y & Y &  &  & Y &  \\ \hline 
38 & Madhusudan et al., 2011 & Quantitative &  &  &  &  & Y & Y &  & Y &  \\ \hline 
39 & Taquechel, 2013 & Quantitative &  &  &  & Y &  & Y &  & Y &  \\ \hline 
40 & Weaver and Marla, 2018 & Quantitative &  & Y &  &  &  &  & Y & Y &  \\ \hline 
41 & CISA MTA, 2023 & Quantitative &  & Y &  &  & Y & Y &  & Y & Y \\ \hline 
42 & Loose et al., 2023 & Quantitative &  &  &  &  & Y &  & Y & Y &  \\ \hline 
43 & Thekdi and Santos, 2016 & Quantitative &  &  &  &  &  & Y &  & Y &  \\ \hline 
44 & Monson et al., 2007 & Quantitative &  &  &  &  &  & Y &  & Y &  \\ \hline 
45 & Li et al., 2021 & Quantitative &  &  &  &  & Y &  &  & Y &  \\ \hline 
46 & Woo et al., 2013 & Quantitative &  & Y &  &  &  &  &  & Y &  \\ \hline 
47 & Molavi et al., 2020 & Quantitative &  &  & Y &  &  &  & Y & Y &  \\ \hline 
48 & Amrou et al., 2019 & Quantitative &  &  &  &  &  &  & Y & Y &  \\ \hline 
49 & Bastug et al., 2020 & Quantitative &  & Y & Y &  &  &  &  & Y &  \\ \hline 
50 & Becker, 2017 & Quantitative &  & Y &  & Y &  &  &  & Y &  \\ \hline 
51 & Yen et al., 2023 & Quantitative &  & Y & Y &  &  &  &  & Y & Y \\ \hline 
52 & Asperen et al., 2011 & Quantitative &  &  &  & Y &  &  & Y & Y &  \\ \hline 
53 & Thorisson et al., 2019 & Quantitative &  &  &  &  &  &  & Y & Y &  \\ \hline 
54 & Wei et al., 2020 & Quantitative &  &  &  & Y &  &  &  & Y &  \\ \hline 
55 & Balakrishnan et al., 2020 & Quantitative &  &  &  &  & Y &  &  & Y &  \\ \hline 
56 & Askin et al., 2024 & Quantitative &  & Y & Y &  &  &  &  & Y & Y \\ \hline 
57 & Koroleva et al., 2019 & Quantitative &  &  & Y &  &  &  & Y & Y &  \\ \hline 
58 & Vanye et al., 2021 & Quantitative &  &  &  &  &  &  &  & Y &  \\

\bottomrule
\end{tabular}
\end{sidewaystable}

\section*{Acknowledgment}

This material is based upon work supported by the U.S. Department of Homeland Security under Grant Award Number 17STQAC00001-07. The views and conclusions contained in this document are those of the authors and should not be interpreted as necessarily representing the official policies, either expressed or implied, of the U.S. Department of Homeland Security.

\section*{Declaration of Interest}

The authors declare that they have no known potential conflict of interest.

\section*{Author Contribution}

All authors critically revised the manuscript for intellectual content, approved the final version to be published, and agree to be fully accountable for all aspects of the work. Digvijay Redekar was responsible for the conceptualization, methodology, and validation of the study, as well as for drafting the original manuscript and creating visualizations. Ronald Askin contributed to the methodology, provided essential resources, conducted validation, and played a key role in the review and editing process; he also offered supervision and secured funding for the research. Feng Ju was involved in the methodology, provided resources, performed validation, and contributed to the review and editing of the manuscript, and supervision throughout the study.

\section*{Data Availability}

The dataset utilized in this paper is available from the corresponding author upon request.

\newpage

\end{document}